\documentclass[12pt]{article}
\usepackage{geometry,float}
\usepackage{amssymb}
\usepackage{amstext,enumerate}
\usepackage{amsmath,amssymb,amsthm,textcomp}
\usepackage{enumitem}
\usepackage{bm}
\usepackage{graphics}
\usepackage{graphicx}
\graphicspath{{img_hires/pngs/}}
\usepackage[export]{adjustbox}
\usepackage{multicol, multirow}
\usepackage[font=small,skip=1pt]{caption}
\usepackage{amsthm}
\usepackage{amsmath}
\usepackage{latexsym}
\usepackage{rotate}
\usepackage{lscape}
\usepackage{color}
\usepackage{epsfig,epsf,psfrag}
\usepackage{sectsty}
\usepackage{epsfig}
\usepackage{multirow}
\usepackage{amsmath}
\usepackage{url}
\usepackage{amsthm}
\usepackage{amssymb}
\usepackage{sectsty}
\usepackage{epsfig,natbib}
\usepackage{rotate}
\usepackage{lscape}
\usepackage{setspace}
\usepackage{subcaption}
\usepackage{float}
\usepackage{mathtools}
\usepackage{booktabs}
\newtheorem{Theorem}{Theorem}%[section]

\sectionfont{\centering\bf\sf\normalsize}
\subsectionfont{\sf\normalsize}

\renewcommand{\baselinestretch}{1.6} %{2} %{1.4}
\newcommand{\single}{\renewcommand{\baselinestretch}{1.2}\normalsize}
\newcommand{\double}{\renewcommand{\baselinestretch}{1.63}\normalsize}

  \renewenvironment{thebibliography}[1]{
    \begin{oldthebibliography}{#1}
      \setlength{\parskip}{0ex}
      \setlength{\itemsep}{0ex}
  }
  {
    \end{oldthebibliography}
  }

\newcommand{\bea}{\begin{eqnarray*}}
\newcommand{\eea}{\end{eqnarray*}}
\newcommand{\be}{\begin{eqnarray}}
\newcommand{\ee}{\end{eqnarray}}
\newcommand{\ed}{\end{document}}

\newcommand{\btab}{\begin{tabular}}
\newcommand{\etab}{\end{tabular}}

\newcommand{\bi}{\begin{itemize}}
\newcommand{\ei}{\end{itemize}}
\newcommand{\bfi}{\begin{figure}}
\newcommand{\efi}{\end{figure}}
\newcommand{\ben}{\begin{enumerate}}
\newcommand{\een}{\end{enumerate}}
\newcommand{\bay}{\begin{array}}
\newcommand{\eay}{\end{array}}

\def\bco{\iffalse}

\newcommand{\no}{\noindent}
\newcommand{\bc}{\begin{center}}
\newcommand{\ec}{\end{center}}

\newcommand{\mt}{\mathcal{T}}

\DeclareMathOperator*{\argmin}{argmin}

\bibpunct{(}{)}{;}{a}{}{,}

\def\references{\bibliography{xc-ref-2}}
\begin{document}

\thispagestyle{empty} 

\single \bc { \Large \bf \sc  Cross-Component Registration for\\ Multivariate Functional Data, \\ \vspace{3mm} With Application to Growth Curves}%with Application to Longitudinal Growth Curves
\footnote{Research supported by NSF Grant DMS-1712864.}

%$^{\dagger *}$}
\vspace{0.5in}

Cody Carroll$^1$,  Hans-Georg M\"uller$^1$, and Alois Kneip$^2$ \\
$^1$ Department of Statistics, University of California, Davis \\
$^2$ Department of Economics, Universit\"at Bonn\\

 \ec \centerline{July 2020}

\vspace{0.4in} \thispagestyle{empty}
\bc{\bf \sf ABSTRACT} \ec \vspace{-.1in} \no 
\setstretch{1}
Multivariate functional data are becoming ubiquitous with advances in modern technology and are substantially more complex than univariate functional data. We propose and study a novel model for multivariate functional data where the component processes are subject to mutual time warping. That is, the component processes exhibit a similar shape but are subject to systematic phase variation across their time domains. To address this previously unconsidered mode of warping, we propose new registration methodology which is based on a shift-warping model. Our method differs from all existing registration methods for functional data in a fundamental way. Namely, instead of focusing on the traditional approach to warping, where one aims to recover individual-specific registration, we focus on shift registration across the components of a multivariate functional data vector on a population-wide level. Our proposed estimates for these shifts are identifiable, enjoy parametric rates of convergence and often have intuitive physical interpretations, all in contrast to traditional curve-specific registration approaches.  We demonstrate the implementation and interpretation of the proposed method by applying our methodology to the Z\"urich Longitudinal Growth data and study its finite sample properties in simulations.

\no {KEY WORDS:\quad Component processes; Functional data analysis; Growth curves; Multivariate functional data; Shift registration; Time warping}.
%\thispagestyle{empty} \vfill
%\noindent \vspace{-.2cm}\rule{\textwidth}{0.5pt}\\
%{\small Matthew Dawson is  Graduate Student Researcher and Hans-Georg M\"uller is Professor at the Department of Statistics, University of California, Davis. 
%This research was  supported by NSF grants DMS-1228369 and
%DMS-1407852. The data used in this paper are from the Alzheimer's Disease Center at University of California Davis, supported by NIH and NIA grant P30 AG10129.
%We are grateful to the Editor, an Associate Editor and two referees for
%constructive critiques, which led to many improvements in the
%paper. Special thanks are due to  Wenwen Tao and Wesley Thompson for discussions of the problem of longitudinal snippet data.}

\newpage
\pagenumbering{arabic} \setcounter{page}{1}

\newpage
\pagenumbering{arabic} \setcounter{page}{1} \double

\section{Introduction}
\label{s:intro}

Multivariate functional data are often encountered in biological or chemical processes that  are continuously measured for a group of subjects or observational units.  Such processes arise in many longitudinal studies, especially in the biomedical sciences, the scopes of which range from human growth to time courses of protein levels during metabolic processes \citep{park:17, mull:05:2}.   With the increasing ubiquity of multivariate functional data, the study of how to treat such data has recently become a very active field, particularly in the context of clustering \citep{brunel2014, jacq:14, park:17}, functional regression \citep{chio:12,chio:16}, and in terms of general modeling of functional data \citep{clae:14}. Common approaches for analyzing multivariate functional data have focused on dimension reduction via multivariate functional principal components (MFPCA) \citep{zhou:08, chio:14, happ:18} or decomposition into component-specific processes and their interactions \citep{chio:16}.

In applications such as growth curves, if  we view multivariate longitudinal data as generated by an underlying $p$-dimensional smooth stochastic process, the component curves of the functional vector may exhibit mutual time warping. If left unchecked, such vector component warping may distort principal components and inflate data variance, while if handled properly, it may yield intuitive physical interpretations and a more parsimonious representation of the data. As far as we know, the idea of explicitly modeling time relations between component processes has not yet been considered for multivariate functional data, which allows one to take advantage of repeated observations of a multivariate process for a cohort of subjects. 

Typically, for each subject in longitudinal studies one has measurements on a grid of time points, where recordings are possibly contaminated with measurement error. Often these measurements are multivariate, notably in growth studies \cite[e.g.,][]{mull:18}, which prompts consideration of functional methods which are geared towards repeatedly sampled multivariate functional data. The Z\"urich Longitudinal Growth Study motivated us to model such multivariate functional data by allowing the components to be mutually time-shifted against each other, as some components of growth may systematically precede others. 

The idea of warping across components is most pragmatic when the component processes of multivariate functional data exhibit similarity in their shapes. In the case of growth studies, each body part's component process follows the same general pattern: a period of rapid development during infancy which then slows to a roughly constant rate of growth until puberty, at which time the growth velocity peaks (i.e., the pubertal growth spurt) before decreasing to zero as the subject reaches adulthood \citep{mull:84}. The multivariate aspect of these growth curves allows us to compare the growth processes of different parts of the body.  For example, it may be that legs undergo their growth spurt earlier in life than arms do. It is an interesting biological question to search for a common process that ordinates the timings of growth spurts across body parts. Another situation where this phenomenon arises is in the above-mentioned recordings of protein levels during metabolic processes. Certain biological functions are associated with peaks and valleys of certain protein levels and their relative timings expose the order of the underlying enzymatic mechanisms at work.

Data from the Z\"urich Longitudinal Growth Study were used previously to investigate the timing of growth spurts across body parts using a phase-clustering model \citep{park:17}. Our study uses the same data but instead emphasizes the investigation of phase variations in the component growth velocity curves to establish time relations. In particular, we investigate mutual time warping in the derivatives across the components of the multivariate functional processes during a growth spurt window, as derivatives are more informative about human growth than the growth curves themselves. Specifically, we assume a model which uses relative time shifts between component processes to establish their pairwise time relations. Information about the relative shifts between pairs of components may then be combined to inform the full system of relative timings across body parts. We emphasize that our approach, while motivated by growth data, is by no means limited to this application and can also be implemented for multivariate functional data which has neither a well-defined time origin nor an endpoint, as in the case of blood protein time courses \cite[e.g.,][]{mull:05:2}. 

The organization of this paper is as follows. Section 2 motivates and establishes a shift-warping model for the cross-component registration problem. In Section 3 and 4 we estimate the proposed model components in the pairwise and general settings, respectively. An application to human growth curves is discussed in Section 5. A simulation study which illustrates the stability of the method even in the presence of nuisance peaks and sizeable measurement error is explored in Section 6. Section 7 contains theoretical results, with accompanying proofs appearing in the appendix. Specifically, we find that under a quadratic curvature assumption, one attains parametric rates of convergence for cross-component shift estimates.

\section{A Shift-Warping Model for Multivariate Time Relations}

\begin{figure}
	\centering
	\includegraphics[width=\linewidth]{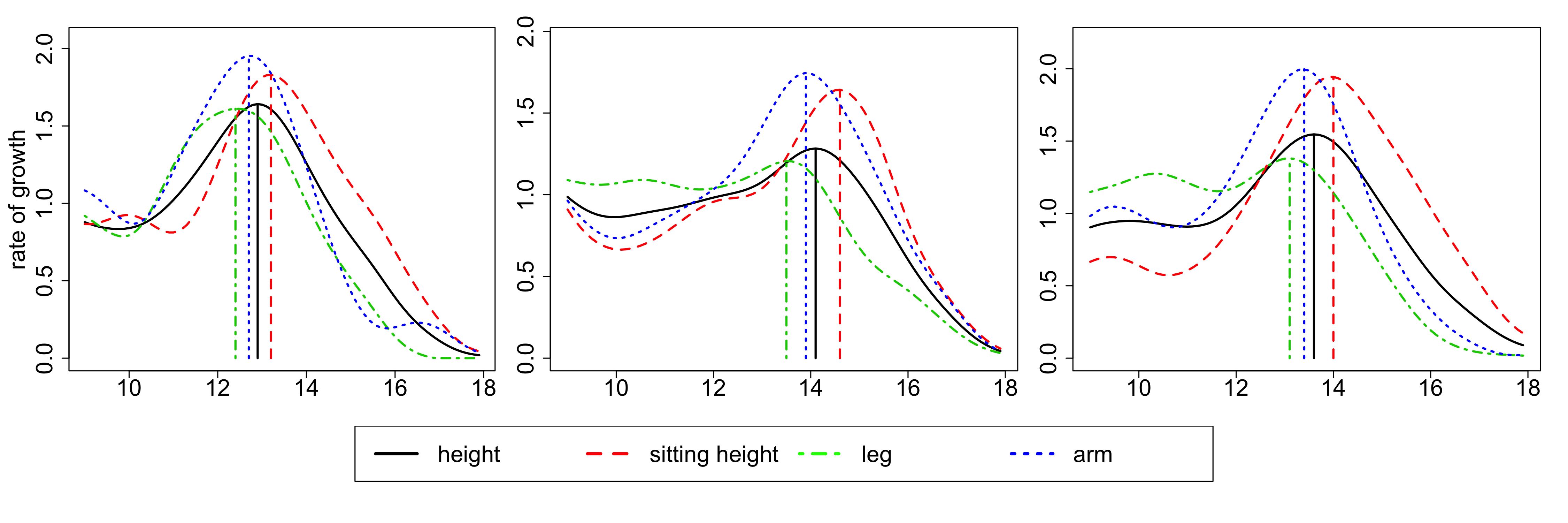}
	\caption{Three children's growth velocities for standing height (black, solid), sitting height (red, dashed), leg length (green, dot-dashed), and arm length (blue, dotted). Peak velocity positions are marked with vertical lines and can be used as rough markers of pubertal onset for each modality. \textit{This figure appears in color in the electronic version of this article, and any mention of color refers to that version.}}
	\label{fig-demo}
\end{figure}

To illustrate the idea of mutual component warping, consider the growth velocities for a handful of representative children in the Z\"urich Longitudinal Study (Fig. \ref{fig-demo}), which will be revisited in its entirety in Section 5. We consider pubertal growth, i.e. growth curves are evaluated in the interval $\mathcal{I}=[9,18]$ ranging from 9 to 18 years.
Each child has four growth velocity curves, each corresponding to a different body part. The peaks represent the moment of maximal rate of growth and can be used as a crude measure of the timing of pubertal growth spurt for that modality. For ease of viewing we mark these locations in time with vertical lines in Figure \ref{fig-demo}. 

A key observation is to recognize that regardless of when the child underwent puberty, the ordering of the spurts is consistent: legs undergo their growth spurts first, then arm length and standing height roughly together, followed by sitting height. This pattern in pubertal spurts was briefly discussed in the descriptive growth studies of \cite{shee:99} and suggests that there is a population-wide mutual component warping occurring across the four modalities. Note also that the time differences between modalities are relatively consistent across children, despite individual differences in the age of the pubertal onset. This is worth highlighting for two reasons: (1) it motivates the estimation of a fixed population-wide set of shift parameters, and (2) it shows that cross-component registration makes sense even in the presence of subject-specific time warping, which is the usual mode of warping considered in univariate functional data. Subject-specific registration is a complex and varied field and we do not attempt to provide a comprehensive review here; for a recent overview of traditional warping methods, we refer less familiar readers to \cite{marr:15} and \cite{wangrev:16}.

To register these curves across components, we propose a shift-warping model, which provides a simple and interpretable method for cross-component alignment of growth data. From a methodological point of view, our approach builds on basic ideas in the literature on parametric and semi-parametric modeling of growth and related phenomena. In applied work on human growth, empirical studies often utilize parametric models  \citep{mila:00}. One of the most popular classes of models has been proposed by \cite{pree:78}; for a recent application, see, e.g. \cite{bani:17}. All of these models make use of shift parameters $\theta_{ij}$ to capture the main differences in individual timings.  For $p$-dimensional multivariate functional data, $\{X_{i1}(t),\ldots,X_{ip}(t)\}^T$,  $i=1,\ldots,n$, which we consider here on a domain $\mathcal{I}$ that covers the pubertal period, an extension of the existing models to the multivariate case is as follows. 

For some function $G$ and some additional parameter vectors $\bm{\xi}_{ij}$ one posits that, with  time shifts $\theta_{ij}$,  the 
growth curve for the $j^{th}$ component of the $i^{th}$ subject has the form 
\begin{equation}
X_{ij}(t) = G(\bm{\xi}_{ij},t-\theta_{ij}),  \quad j=1,\dots,p, \,\, i=1, \ldots,n,\,\, t\in \mathcal{I},
\label{eq-1.1}
\end{equation}
where previously only cases with $p=1$ have been considered. As fully parametric specifications were found to lack accuracy, various semi-parametric extensions have been proposed for the one-dimensional case.  For example, for standing height, in the case $p=1$, \cite{knei:95} assumed a shape-invariant model with $G(\bm{\xi}_{ij},t-\theta_{ij})=\xi_{ij;2} f\{\xi_{ij;1}(t-\theta_{ij})\}+\xi_{ij;3}$ for real-valued parameters $\xi_{ij;1},\xi_{ij;2},\xi_{ij;3}$, and an unknown real-valued function $f$ which is estimated from the data. The $k$-mean alignment introduced by \cite{sang:10} may be seen as a generalization of this framework, where it is assumed that the population can be decomposed into $K$ disjoint clusters, and 
individual functions belonging to each cluster can be approximately described by a shape-invariant model with respect to a cluster-specific template function $f_g$, $g\in\{1,\dots,K\}$. 

In the following we assume that growth data follow a multivariate and flexible version of models of type \eqref{eq-1.1}, under the natural assumption that the shift parameters $\theta_{ij}$ can be decomposed in the form $\theta_{ij}=\theta_{i}+\theta_{j}$, where $\theta_i$ is specific for the individual, while $\theta_j$ is specific for the component. Then \eqref{eq-1.1} may be rewritten in the form
\begin{equation}
X_{ij}(t) = G(\bm{\xi}_{ij},t-\theta_{i}-\theta_j)  \equiv G^*(\bm{\xi}_{ij}^*,t-\theta_j),  \quad\text{ where } \bm{\xi}_{ij}^*=( \bm{\xi}_{ij},\theta_i).
\label{eq-1.2}
\end{equation}

Motivating our alignment procedure is that, for a given individual $i$, the
component functions $X_{i1}(t),\ldots,X_{ip}(t)$ can be made more similar when removing the different shift parameters $\theta_1,\dots,\theta_p$. The most favorable situation arises if shifts constitute the only important difference between components such that $\bm{\xi}_{ij}\equiv \bm{\xi}_{i}$ is independent of $j=1,\dots,p$. Then with $Z_i(s):=G^*(\bm{\xi}_{i}^*,s)$ we arrive at
\begin{equation}
X_{ij}(t) = Z_i(t-\theta_j),  \quad j=1,\dots,p, \,\, i=1, \ldots,n,\,\, t\in \mathcal{I},
\label{eq-1.3}
\end{equation}
so that 
$E\int_\mathcal{I}\{X_{ij}(t+\theta_j)-X_{il}(t+\theta_l)\}^2dt= E\int_\mathcal{I}\{Z_i(t)-Z_i(t)\}^2dt=0$ for all $j,l\in\{1,\dots,p\}$; to apply this argument  will require some pre-processing in order to eliminate scale differences between the different components (see Section 5).

In the context of growth curves, subject-specific alignment based on nonparametric monotonic warping functions $h_i: \mathcal{I} \to \mathcal{I}$ has been studied extensively \citep{gass:90, knei:95, gerv:04, mull:08:6}. Higher dimensional problems of subject-specific registration have been considered through  the lens of elastic shape analysis \citep{sriv:10,sriv:16}, or reduced to the problem of aligning a univariate curve generated from the component curves \citep{rams:14}. It can be seen from \eqref{eq-1.2} and \eqref{eq-1.3} that in our context such functions $h_i$ do not play any role and may simply be part of the parameter set $\bm{\xi}_i$. We therefore emphasize that in the non-traditional warping framework presented here, the pertinent issues are fundamentally different from those considered in the subject-specific warping framework discussed in the cited articles. In short, that it bypasses dealing with individual warping functions is a strength of our method and allows us to side-step the identifiability problems associated with subject-specific registration. A more detailed discussion of this matter in the context of the Z\"urich data  can be found in the Supplement.

It is especially noteworthy that we obtain a $\sqrt{n}$-rate of convergence for the estimated time shifts to their targets under mild regularity conditions (see Section 7). Such fast convergence rates cannot be obtained in traditional warping approaches, since these focus on individual warps rather than component-specific warping and therefore require identification of $n$ time alignments, where $n$ is the sample size, whereas in our approach there are only $p$ components that need to be considered, where $p$ is the fixed dimension of the multivariate process. Of course, in some circumstances the model in \eqref{eq-1.3} may just serve as a convenient approximation of a more nuanced warping relation between components. We discuss the potential for continuous analogues of cross-component shift-warping techniques in the Concluding Remarks.

A further distinction between cross-component warping as proposed here and the common subject-specific approach is that the latter traditionally views the presence of individual warping functions as a nuisance characteristic of the data to be accounted for in order to correctly analyze underlying functional features of interest; for example, curves will be registered first before conducting a functional principal component analysis (FPCA). In contrast, we argue that investigation of cross-component warping and the shift parameters $\theta_1,...,\theta_p$ provide insight into inter-component relationships and, when applicable, are an essential aspect of multivariate functional data that is of genuine interest rather than a nuisance.

\section{Bivariate Cross-Component Registration}
\subsection{Pairwise-Shift Estimation}
We introduce here the main idea of registering different component times across modalities, which we call Cross-Component Registration (XCR). As explained in the previous section, XCR differs in key aspects from traditional warping, which is also known as curve registration or alignment \citep{rams:05, knei:92, silv:95}, as it aims at a situation where the component curves of a multivariate functional process are time-shifted versions of one another. A major difference is that instead of estimating $n$ individual warping functions, which align curves across subjects and the determination of which is the goal of traditional curve warping methods, our new approach targets a $p$-vector of shift parameters for the case of $p$-dimensional functional data. These component-wise shifts are then applied uniformly across all subjects to mutually align the component curves.

In  the following, we write $(X_1,\ldots,X_p)^T$ to represent the generic underlying multivariate process and $\{X_{i1}(t),\ldots,X_{ip}(t)\}^T, \,\, i=1,\ldots,n$,   for a sample of realizations of the functional vector. One may assume \textit{a priori} smoothness of curves or may preprocess the data with a smoothing method if the curves are subject to measurement error.  In this subsection we consider the case of multivariate functional data with just $p=2$ component curves to introduce the main ideas, and will then discuss the extension to $p>2$. To fix the idea, consider a sample of bivariate functional processes, writing $\{X_{i1}(t),X_{i2}(t)\}_{i=1}^n$ for the observed i.i.d. realizations of the bivariate process $(X_1,X_2)$, and assume that the domain of both component processes is a compact interval $\mathcal{T}=[0,T]$. As a criterion for alignment and to determine the optimal shift, we  aim to minimize the $\mathcal{L}^2$-distance between functions on a subinterval $\mathcal{I} \subset \mt$; see the discussion below. Using a simple shift-warp family under the $\mathcal{L}^2$-norm allows for a straightforward and clear interpretation of the relationship between two components and has been used previously in the context of shape-invariant modeling  \citep{hard:90, knei:95, silv:95}.

\par Specifically, we aim for the optimal value of the parameter $\tau$, the pairwise {\it cross-component (XC) shift} as the minimizer of
\begin{equation}
\Lambda(\tau)=E\int_\mathcal{I} \{X_{1}(t)-X_{2}(t-\tau)\}^2dt,
\label{eq-2.1.2}
\end{equation}
with associated sample version
\begin{equation}
L_n(\tau)=\frac{1}{n}\sum_{i=1}^n\int_\mathcal{I} \{X_{i1}(t)-X_{i2}(t-\tau)\}^2dt
\label{eq-2.1.3}
\end{equation}
and sample-based shift parameter estimate
\begin{equation}
\widehat{\tau}=\argmin_\tau L_n(\tau),
\label{eq-2.1.4}
\end{equation}
targeting

$\tau_0=\argmin_\tau \Lambda(\tau).$

Integrating over a subinterval $\mathcal{I}$ rather than the whole interval is a device that is necessary in order to ensure that both the shifted and unshifted curves are defined on the domain of integration. If we did not specify a suitable subinterval $\mathcal{I} \subset [0,T]$ that stays away from both 0 and $T$,  shifting a curve forward or backward may result in a subinterval of integration in which one of the curves is defined while the other is not, making it impossible to compute their $\mathcal{L}^2$-distance. To be precise, we partition the data domain $\mathcal{T}$ into three disjoint intervals $\mathcal{T} = R_1 \cup \mathcal{I} \cup R_2$, where $\mathcal{I}=[r_1,r_2]$ is the subinterval of integration and $R_1=[0,r_1)$ and $R_2=(r_2, T]$ are the remaining intervals on the boundary. Note that this partitioning implies that the magnitude of pairwise shift estimates cannot exceed the length of the relevant remainder interval, depending on the direction of the shift. This subtlety suggests that the choice of subinterval of integration $\mathcal{I}$ is not trivial and should be done  carefully and data-adaptively.

\subsection{Subinterval Selection}
We propose the following guidelines for subinterval selection:  $\mathcal{I}$ should be chosen to (1) include the critical features of the sample curves, and (2)  avoid censoring estimates of pairwise shifts. For example, in our application to the Z\"urich data, we choose $\mathcal{I}$ to range from the earliest age of pubertal onset to the age of adulthood. Doing so ensures the inclusion of the main pubertal growth spurt peaks which are the structural features to be aligned across components \citep{gass:95}. Unreasonable estimates may occur if the subinterval is too small, as an inappropriately narrow window may discard the features to be aligned for a subset of individuals. 

The problem of subinterval selection was discussed previously  in \cite{knei:95} and we follow their convention to seek an ``overlapping interval" across all individuals, described as follows. Individual intervals $\mathcal{J}_i$ are chosen such that information about structural landmarks for the $i^{th}$ individual are contained entirely in $\mathcal{J}_i$. Then the overlapping interval $\mathcal{J}$ is defined as $\mathcal{J} = \cup_{i}\mathcal{J}_i$ and guarantees that all individuals' structural features are included. One can then either simply use this overlapping interval as the subinterval of integration, i.e., let $\mathcal{I}=\mathcal{J}$,  or choose $\mathcal{I}$ such that $\mathcal{J}\subset\mathcal{I}$ and $\mathcal{I}$ has some relevant physical meaning. An example for the latter case is demonstrated  in the data application of Section 5.

In the more general setting with more than two components, we will encounter several pairwise time shifts between sets of component curves. To distinguish between these, we write $\tau_{jk}$ to denote the relative time shift which moves component $k$ to component $j$. Note that the sample and population time shifts are symmetric in the sense that $\tau_{jk}=-\tau_{kj}$. The problem of estimating general cross-component shift parameters $\theta_1,...,\theta_p$ can be solved after the estimation of all of the pairwise shift parameters $\tau_{jk}$ for $1\leq j<k \leq p $, as discussed in the following section.

\section{General Cross-Component Registration}

We now extend the methodology for bivariate cross-component registration (XCR)  to the case of $p$-dimensional multivariate functional processes, aiming to align more than two component functions. Assume we observe $p$-variate functional data $\{X_{i1}(t),\ldots,X_{ip}(t)\}^T$ for $i=1,\ldots,n,$ now with $p>2$. We search for a vector of global XC shifts, $\bm{\theta}=(\theta_1,\dots,\theta_p)$, such that when each modality $X_j(t),~j=1,\dots,p,$ is shifted by $\theta_j$, all $p$ curves are aligned. Here it is useful to introduce the idea of an underlying \textit{latent process}, which may be seen as the $Z_i$ component in model (\ref{eq-1.3}).

\begin{figure}
	\centering
	\includegraphics[width=0.99\linewidth]{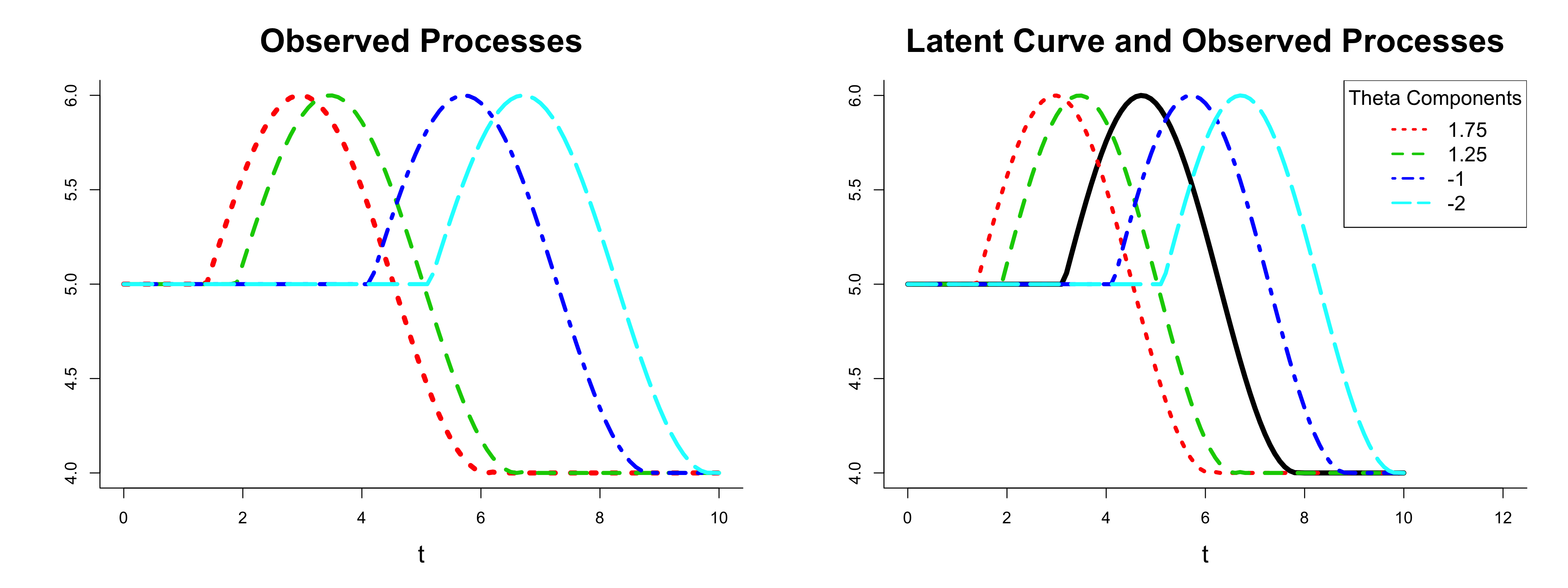}\par
	\caption{Observed components (dashed, left) and latent curve (solid, right) defined by identifiability constraint. \textit{This figure appears in color in the electronic version of this article.}}
	\label{fig-latent}
\end{figure}

To fix the idea, consider  only a single observation of simulated multivariate functional data where the components of the multivariate process are  just time-shifted replicates. Figure \ref{fig-latent} illustrates an example for  $p=4$. A simple approach would be to align the component curves  by fixing one component curve and shifting the others via bivariate XCR to align them with the selected component. However, a major problem with this approach  is that the resulting XC shifts depend on the choice of the fixed component.

These problems can be overcome by assuming that each curve is a shifted version of an unobserved and unshifted latent component, visualized as the solid curve in Figure \ref{fig-latent}. The observed components are then time-shifted with respect to this latent component and the shifts are subject to the constraint $\sum_{j=1}^{p}\theta_j=0$, so that there is no net XC shift from the latent component curve. This assumption is necessary for the identifiability of the shift estimates.

A key observation is that there is a linear relationship between pairwise XC shifts, $\tau_{jk}$, and the global XC shifts, $\theta_j$ and $\theta_k$. Specifically, the pairwise shifts can be expressed as the difference of two global shifts as shown in Eq.~\ref{eq-2.2.0}. Thus, after estimating bivariate XC shifts $\tau_{jk}$ between component functions, we can infer the global XC vector $\bm\theta$, and importantly, the linear map between the two is invariant with respect to the choice of the latent process. 

More explicitly, the linear map $L$ is given by:
\begin{equation}
\tau_{jk}=\theta_j-\theta_k, \quad j,~k =1,\ldots, p, \quad  j < k
\label{eq-2.2.0}
\end{equation}
with constraint $\sum_{j=1}^p \theta_j=0$, so that
\begin{equation}
\bm{\tau}^*=L(\bm{\theta}) =\mathbf{A} \bm{\theta},
\label{eq-2.2.1}
\end{equation}
where $\bm{\tau}^*=(\bm{\tau}^T,0)^T=(\tau_{12},\tau_{13},\ldots,\tau_{(p-1)p},0)^T$ is the pairwise shift parameter vector stacked with 0, $\bm{\theta}=(\theta_1,\ldots, \theta_p)^T$ is the global shift vector of each component function with respect to the latent process, and $\mathbf{A}$ is the matrix of the linear map which corresponds to the contrasts in (\ref{eq-2.2.1}).
Note that $\mathbf{A}$ is of dimension $(p(p-1)/2)\times p$, and is always of full column rank. Explicitly, we write
\begin{equation*}
\resizebox*{!}{0.5\vsize}{
	$\mathbf{A}=\left(\begin{array}{cccccccc}
	1 & -1 & 0 & 0 & 0  & \ldots & 0 & 0\\
	1 & 0 & -1 & 0 & 0  &\ldots  & 0& 0\\
	1 & 0 & 0 &  -1 & 0  &\ldots & 0 & 0\\
	\vdots & \vdots & \vdots & \vdots& \vdots&  \ddots & \vdots & \vdots\\
	1 & 0 & 0 &  0 & 0 &\ldots & 0 & -1\\
	0 & 1 & -1 &  0 & 0  &\ldots  & 0 & 0\\
	\vdots & \vdots & \vdots & \vdots &  \vdots&  \ddots & \vdots & \vdots\\
	0 & 0 & 0 &  0  & 0 &\ldots  & 1 & -1\\
	1 & 1 & 1 & 1  & 1 & \ldots & 1 & 1
	\end{array}
	\right)$}.
\end{equation*}

To implement this approach, we must first estimate the stacked vector of bivariate XC shifts,
$ 
\widehat{\bm{\tau}}^*=(\widehat{\bm{\tau}}^T,0)^T=(\widehat{\tau}_{12},\widehat{\tau}_{13},\ldots,\widehat{\tau}_{(p-1)p},0)^T, 
$ 
leading to the model
\begin{equation}
\widehat{\bm{\tau}}^*=\mathbf{A}\bm{\theta}+\varepsilon, 
\label{eq-2.2.2}
\end{equation}
where $\varepsilon$ is a vector of random noise with mean 0 and finite variance. Once the pairwise shifts  $\widehat{\tau}_{jk}$ are obtained, global shifts $\bm{\theta}$ can be estimated as
\begin{equation}
\widehat{\bm{\theta}}=(\mathbf{A}^T\mathbf{A})^{-1}\mathbf{A}^T\widehat{\bm{\tau}}^*
\label{eq-2.2.3}
\end{equation}
by ordinary least squares. The $p$ component curves will then be aligned (to the latent curve) once they are time-shifted with their respective estimated global XC shifts, $\widehat{\bm{\theta}}$, i.e.,  $X_{ij}(t+\hat{\theta}_{j})$ for $j=1,\dots,p$.

\section{Application to the Z\"urich Longitudinal Growth Study}

\begin{figure}[h!]
	\centering
	
	\includegraphics[width=.75\linewidth]{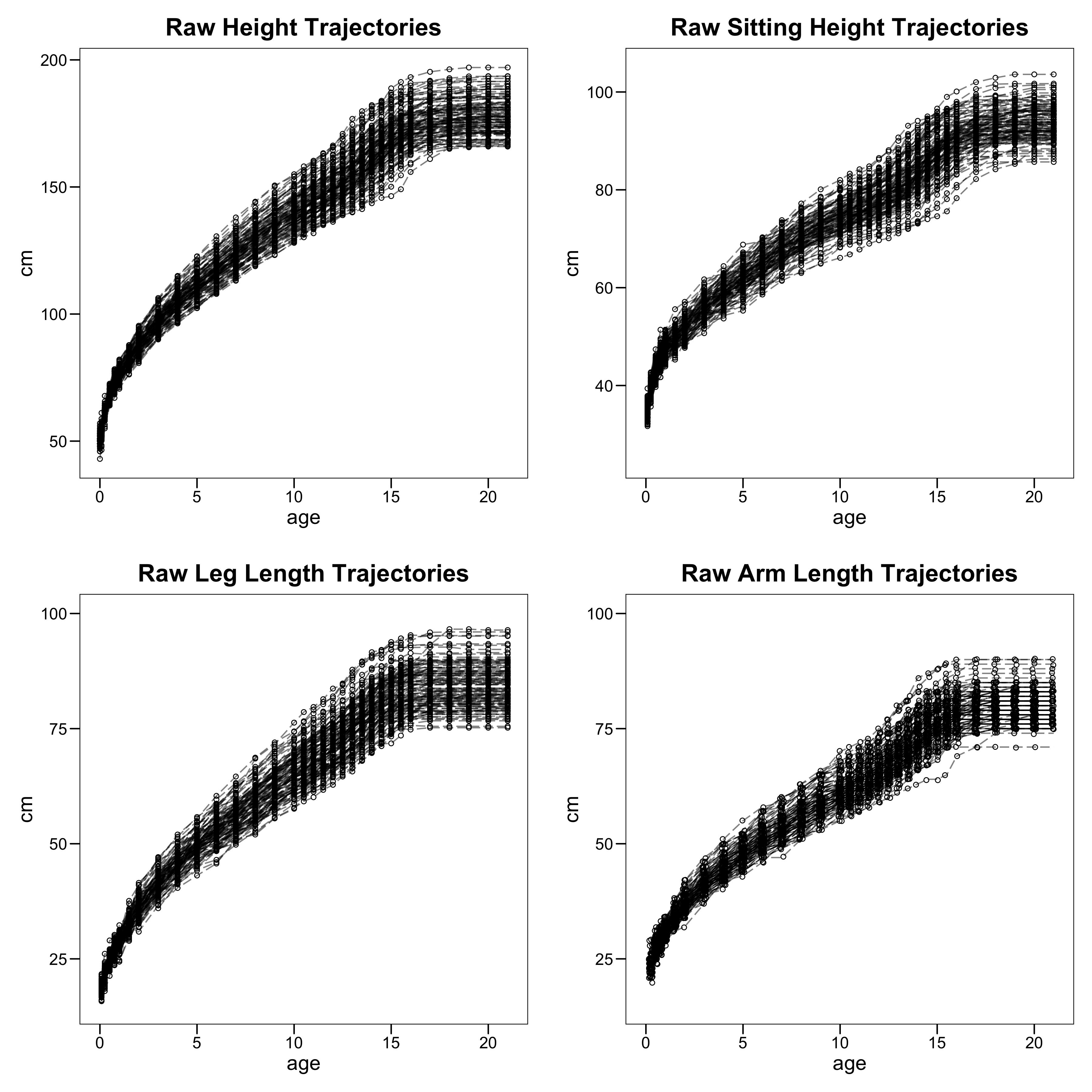}\\
	
	\caption{Raw growth trajectories for all Z\"urich boys.}
	\label{fig-raw}
\end{figure}

From 1954 to 1978, a longitudinal study on human growth and development was conducted at the University Children's Hospital in Z\"urich. Modalities of growth that were longitudinally measured on a dense regular time grid include standing height, sitting height, arm length, and leg length, so that the resulting data can be naturally viewed as multivariate functional data \citep{mull:84:5, knei:89}. The raw trajectories of the $p=4$ component processes for the  children measured are displayed in Figure \ref{fig-raw}, which also indicates the measurement grid. Component curves are initially observed on the domain $\mathcal{T} = [0,20]$,  which can be artificially extended to the right by assuming measurements stay constant in adulthood, since almost all subjects reach full maturation before age 20.

We are especially interested in the timing of pubertal growth spurts, which occur for all individuals between ages 9 and 18 typically. We are using this time  window as the subinterval of integration, $\mathcal{I}$, in accordance with the guidelines of Section 3. A common way to study growth velocities is to examine the derivatives of the growth curves instead of the curves themselves \citep{mull:84}.  The growth velocities have  a peak during puberty, with the apex representing the instant when an individual's growth rate is at its maximum. Previous analysis of human growth curves indicates that there is a difference in the ways that boys and girls undergo puberty \citep{mull:84, eibe:05}. For example, it is widely known that girls begin puberty at younger ages than boys do on average. Accordingly, for the subsequent analysis we separate boys and girls and for brevity display only the results for boys. We estimate the growth velocities, i.e., the derivatives of the growth trajectories, via local weighted linear smoothing using the package \verb|fdapace| \citep{carr:20}.

\begin{figure}[h!]
	\centering
	\includegraphics[width=.75\linewidth]{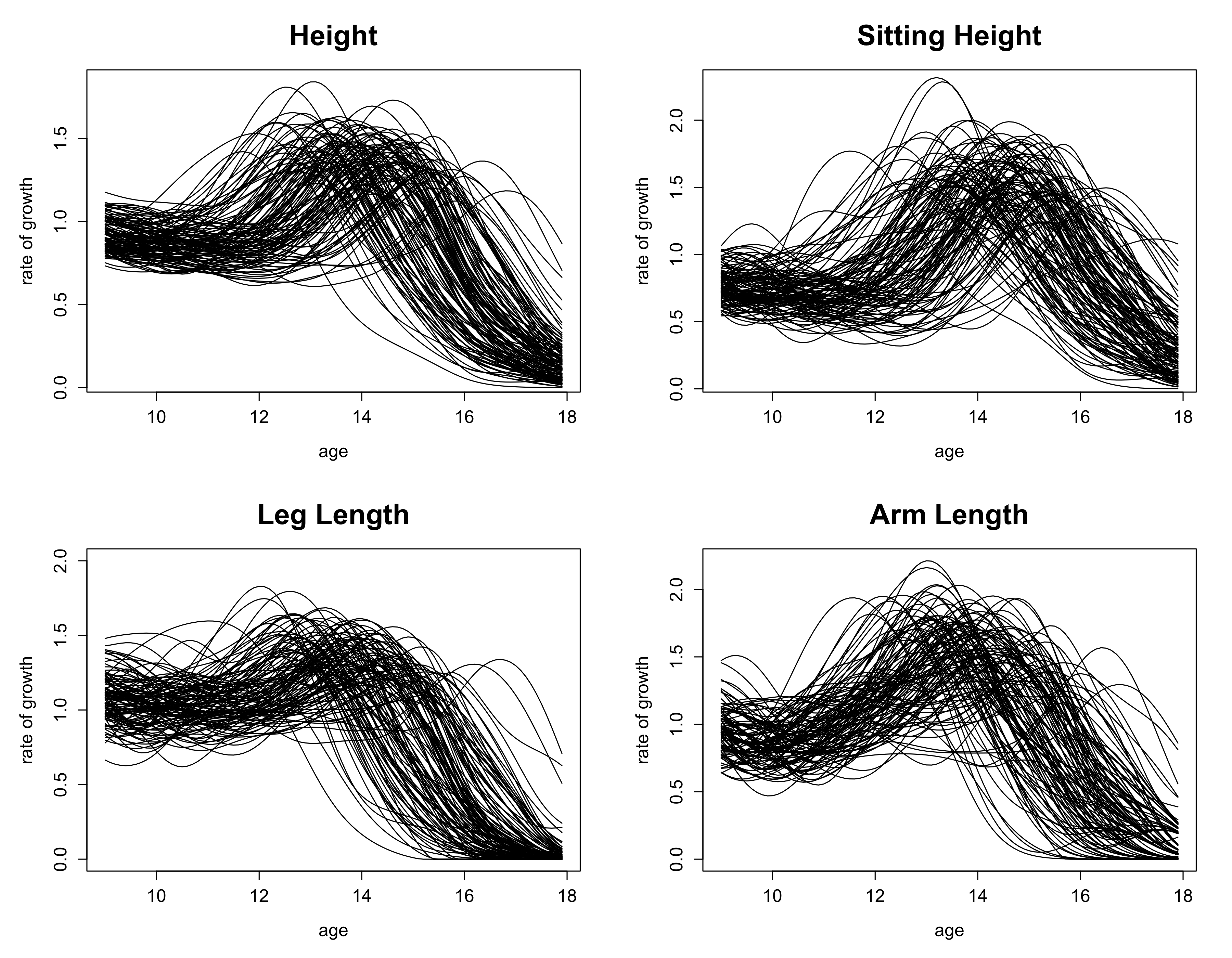}\par
	\caption{Scaled growth velocity curves for Z\"urich boys.}
	\label{fig-scaledvel}
\end{figure}

Because different body parts have different physical sizes, their velocities are also on different scales. We eliminate the majority of this amplitude variation by dividing each function by the total area under the curve, resulting in ``relative velocities" for each modality. Relative velocities have been previously used  in the growth curve literature (see, e.g. \cite{shee:99}) and allow for the comparison of modalities which are on dissimilar scales.  Figure \ref{fig-scaledvel} shows the rescaled derivative estimates for the four growth processes that we consider. After this pre-processing, we now have multivariate functional data with component functions such as those shown for the individuals in Figure \ref{fig-demo}. When we apply the proposed shift model to the growth velocities of the four growth modalities of the Z\"urich data, we obtain  the following estimated global XC shifts (Table~\ref{tab-1}):
\vfill
\newpage

\begin{table}[h!]
	\centering
	\begin{tabular}{ccc}
		\hline
		Component & Modality & Estimate\\
		\hline
		$\theta_1$ & Height & -0.0875\\
		$\theta_2$ & Sitting Height & -0.5850\\
		$\theta_3$ & Leg Length & \hphantom{-}0.5825\\	
		$\theta_4$ & Arm Length & \hphantom{-}0.0900\\
	\end{tabular}
	\vspace{1cm}
	\caption {Estimated global XC shifts for Z\"urich boys. These estimates imply the following ordering of growth spurts: (1) leg, (2) height, (3) arm, (4) sitting height.}
	\label{tab-1}
\end{table}

One can interpret these shift parameters in a pairwise manner.  For example, legs tend to undergo their growth spurts roughly half a year before arms do \break $(\widehat{\tau}_{34}=\widehat{\theta}_3-\widehat{\theta}_4\approx0.5)$ and sitting height trails roughly half a year behind standing height $(\widehat{\tau}_{21}=\widehat{\theta}_2-\widehat{\theta}_1\approx-0.5)$. Our shift estimates and their implied order of growth spurts is consistent with what is known about human growth patterns, as discussed in the descriptive longitudinal studies of \cite{came:82} and \cite{shee:99}.

\vfill
\newpage

\begin{figure}[h!]
	\centering
	\includegraphics[width=\linewidth]{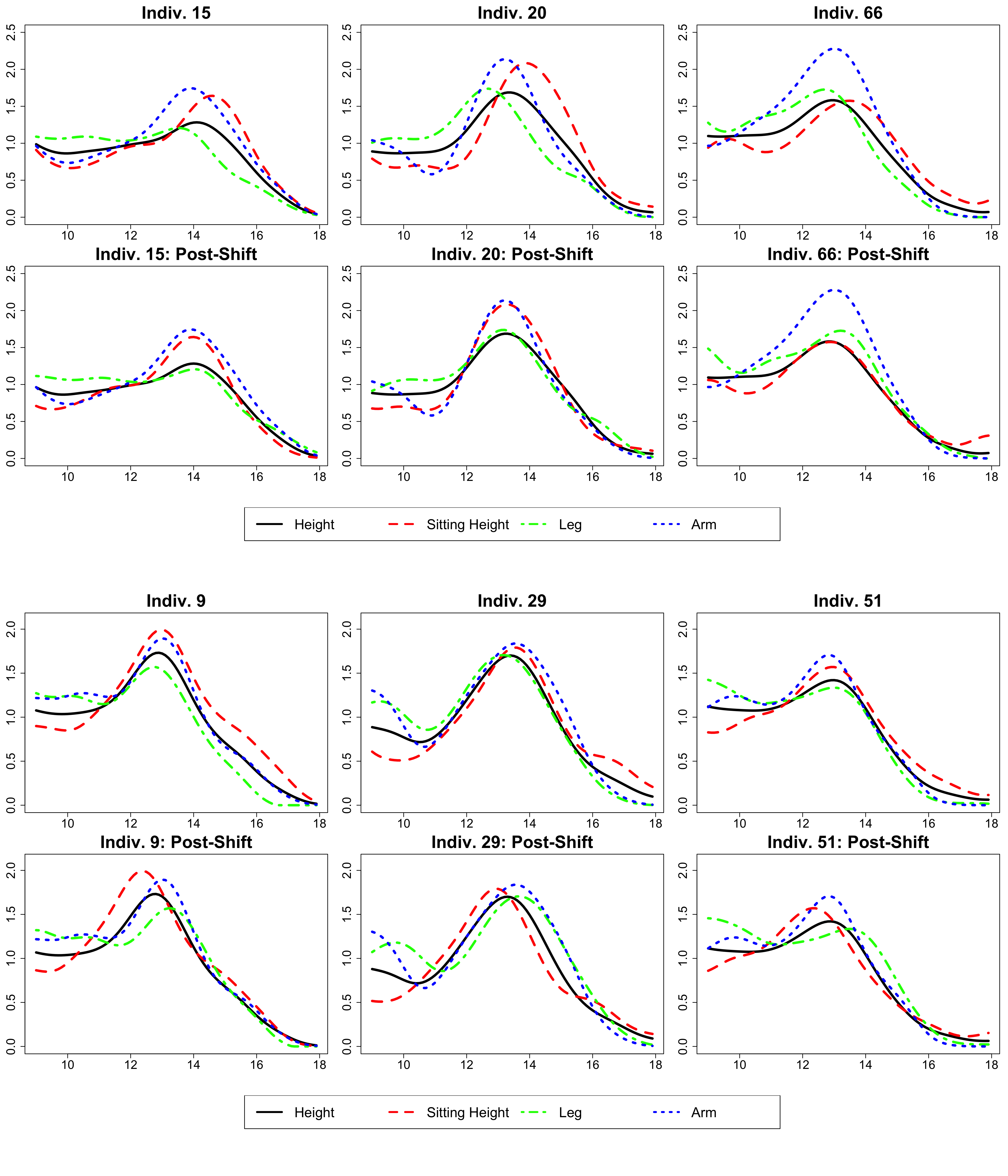}
	\caption{Well-aligned (top) and poorly aligned individuals (bottom) after component alignment.  Growth modalities are standing height (black, solid), sitting height (red, dashed), leg length (green, dot-dashed), and arm length (blue, dotted). \textit{This figure appears in color in the electronic version of this article, and any mention of color refers to that version.}}
	\label{fig-goodbad}
\end{figure}

We next investigate some individuals before and after component alignment for a demonstration of how XC alignment affects the curves. Figure \ref{fig-goodbad} (top) shows three individuals who are representative of the ``average" ordering of growth spurts across modalities, whereas Figure \ref{fig-goodbad} (bottom) displays those who generally went through pubertal spurts for whom the different body parts were already in sync before alignment.
Individuals like those shown in Figure \ref{fig-goodbad} (bottom) for whom alignment moved component curves further away from each other were very rare, and it was common for most individuals to have reduced $\mathcal{L}^2$-distance between the component curves after alignment.

\begin{figure}
	\centering
	\includegraphics[width=.65\linewidth]{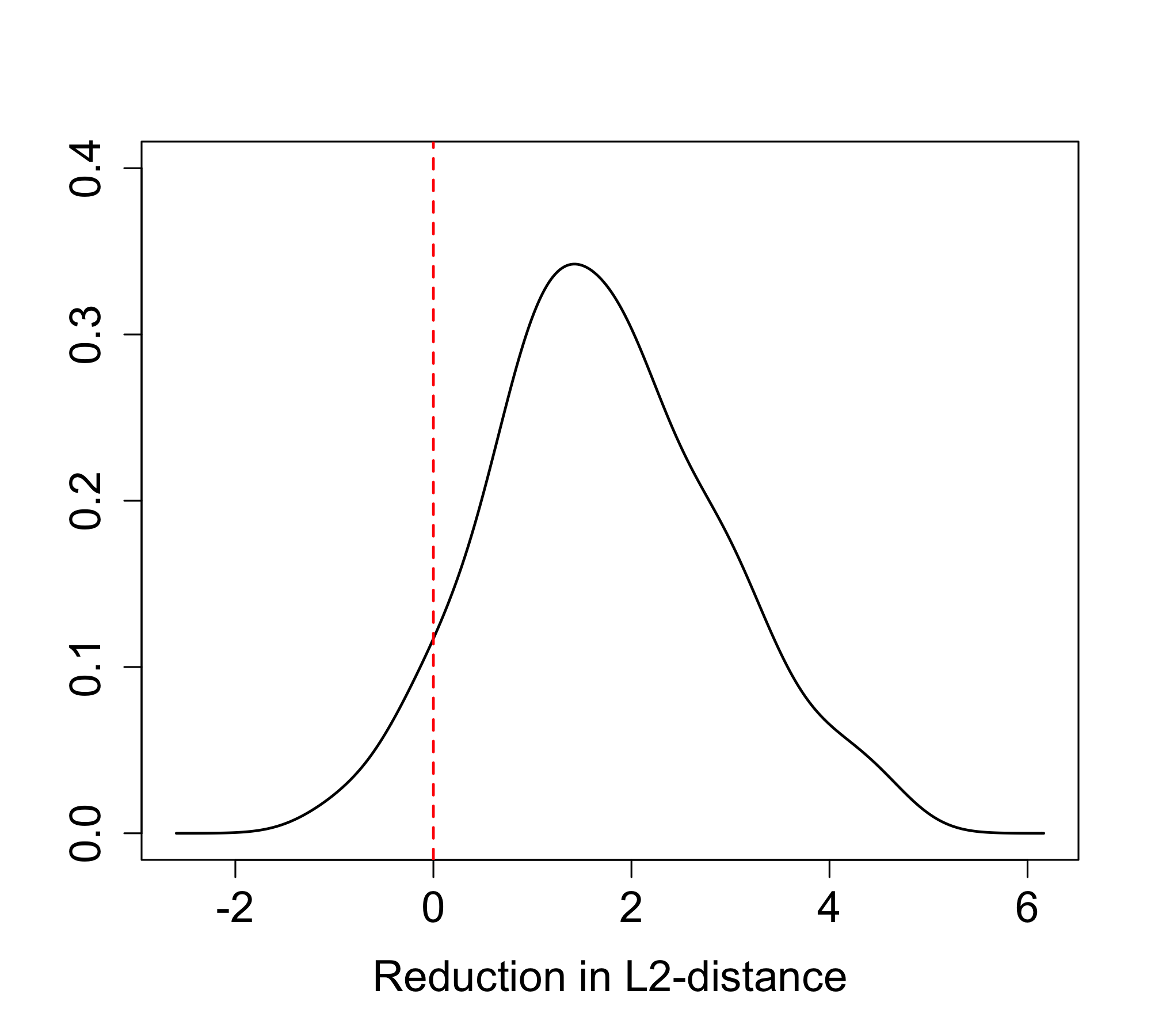}
	\caption{Kernel density estimate of the decrease in total $\mathcal{L}^2$-distance after performing XCR. The dashed line indicates no change.}
	\label{fig-l2reduce}
\end{figure}

To illustrate this further, we use the total cross-component $\mathcal{L}^2$-distance (XD) for an individual as a function of $\bm{\theta}$,
\begin{equation}
\text{XD}_i(\bm{\theta})=\sum_{j<k}\int_\mathcal{I} \{X_{ij}(t+\theta_j)-X_{ik}(t+\theta_k)\}^2dt,
\end{equation}
noting that under a perfect model fit we would have $X_{ij}(t + \theta_j) = Z_i(t)$ for all $j = 1,\dots,p$, and $\text{XD}_i(\bm{\theta}) = \sum_{j<k}\int_\mathcal{I} \{Z_i(t)-Z_i(t)\}^2 dt= 0$.
Figure \ref{fig-l2reduce} displays the distribution of the difference in total  cross-component $\mathcal{L}^2$-distance before and after shifting,  i.e., XD$_i(\bm{0})-$XD$_i(\widehat{\bm{\theta}})$. Here it is noteworthy that implementing component alignment reduced
total $\mathcal{L}^2$-distance in the sample by about $40\%$.
\vfill
\newpage

\section{Simulation Study}

We demonstrate here the superior fit of curves aligned by cross-component registration prior to analysis through FPCA. We use the same base curve $Z(t)=20-.5t+30e^{-\frac{(t-25)^2}{72}}$ on $t\in\mathcal{T}=[0,50]$ as the underlying process dictating the common shape of the component curves and set $\bm{\theta}=(-5,-2.5,2.5,5)$ and $\mathcal{I}=[10,40]$. We contaminate the curves with functional noise, measurement error, and noisy shift parameters by generating contaminated component curves
\begin{equation}
X_{ij}(t_k)=X_{j}(t_k-\theta_j+\eta_{ij})+\zeta_{ij}\sin(\frac{\pi t_k}{5}) + e_{ijk},
\end{equation}
where
$\eta_{ij}\overset{iid}{\sim}\mathcal{N}(0,\sigma_\eta^2)$,
$\zeta_{ij}\overset{iid}{\sim}\mathcal{N}(0,\sigma_\zeta^2)$, $e_{ijk}\overset{iid}{\sim}\mathcal{N}(0,1)$, and $k$ indexes the points on the data grid spanning $\mathcal{T}$ by increments of 0.5. Here the noise on the time domain is introduced through $\eta_{ij}$, while noise on the functional domain is controlled through $\zeta_{ij}$ and $e_{ijk}$,  which correspond to a random amplitude sine wave and minor additive measurement error, respectively.

One can consider each of the component curves as a single noisy warped realization of the underlying latent curve $Z$. We may try to estimate the latent curve by viewing all the component processes for all subjects as a noisy sample of $Z$ and then analyzing them through an established method such as  FPCA. We expect that  failing to account for the component warping will inflate  variances and result in a suboptimal fit, since the cross-component warping masks the features of $Z$, and this is indeed what the following simulations show. 
A sample of $N=100$ curves were fit via FPCA using the first two eigenfunctions, both with and without incorporating XCR. When incorporating XCR, curves were first generated and used to estimate XC shifts, whereupon components were shifted according to these estimates, followed by an FPCA step applied to the thus aligned curves. The first two eigenfunctions were used to fit the sample of aligned curves, and after this fitting step the curves were shifted back to their original domains through the estimated shifts. To quantify the advantage of incorporating XCR, we obtained the integrated mean squared error for both approaches. The benefit of including XCR for various noise scenarios was  measured through the percent decrease in integrated mean squared error for the sample. 

This process was performed $B=1000$ times under low, medium, and high functional noise settings ($\sigma^2_{\zeta}=25,~64,~100$), while letting the noise on the time domain start low and increase until it was on the same scale as the shifts themselves. Table~\ref{tab:sim2} shows the average percent decreases across replications for various settings. The improvements in fit are relatively consistent across functional noise levels. It is noteworthy to observe that once the noise on the domain becomes comparable to that of the shifts themselves (i.e. $\sigma^2_\eta>0.5$), the advantage of XCR starts to decrease. It conforms with expectations that when the within-subject time ordering is highly noise-contaminated, the benefits of performing XCR are lost. At such high shift noise levels there would be little incentive to perform XCR, as exploratory data inspection would not likely indicate the presence of any systematic cross-component warping. 

\begin{table}
	\centering
	\begin{tabular}{|c|c|c|c|c|c|}
		\hline
		Noise Level & $\sigma^2_\eta=0.1$  & $\sigma^2_\eta=0.25$  & $\sigma^2_\eta=0.5$ &$\sigma^2_\eta=1$  &$\sigma^2_\eta=2$  \\
		\hline
		$\sigma^2_{\zeta}=25$ &48.38 & 50.92 &51.39 &44.11 & 16.17 \\
		$\sigma^2_{\zeta}=64$ &48.17 & 50.81 &51.46 &43.75 & 16.64 \\
		$\sigma^2_{\zeta}=100$ & 48.06 & 50.72& 51.37 &43.87 & 16.42 \\
		\hline
		
	\end{tabular}
	\vspace{1cm}
	\caption{Average percent decrease in IMSE after implementing XCR at various levels of contamination on the time ($\sigma_\eta^2$) and functional ($\sigma_\zeta^2$) domains. }
	\label{tab:sim2}
\end{table}

\begin{figure}
	\centering
	\includegraphics[width=\linewidth]{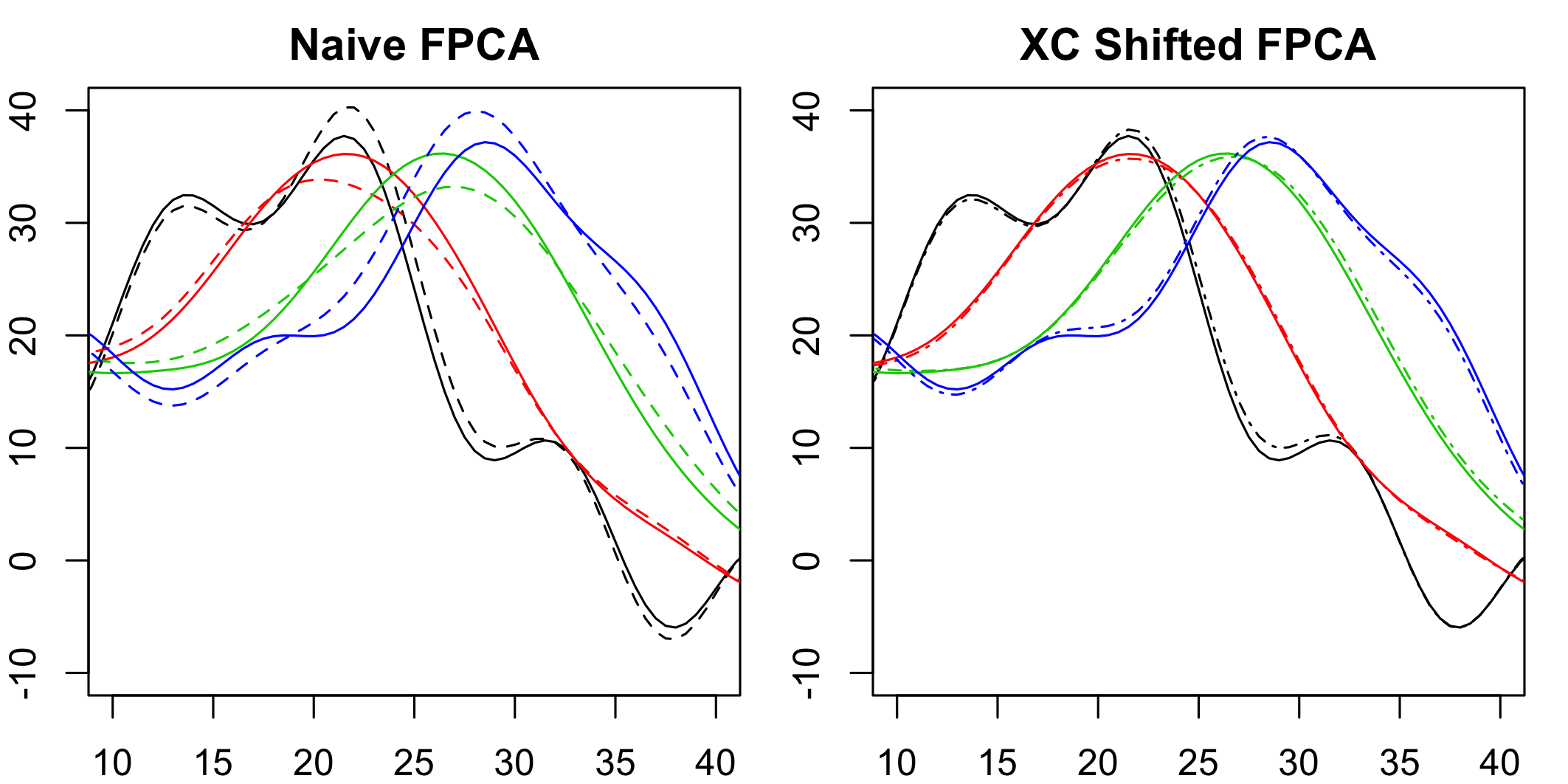}
	\caption{Differences in fit when using just naive FPCA (dashed, left) and FPCA and XCR together (dot-dashed, right), when $\sigma^2_\eta=0.25,~\sigma^2_\zeta=25$. Solid lines represent the original data. \textit{This figure appears in color in the electronic version of this article.}}
	\label{fig:fit}
\end{figure}

A visual comparison of performance for the two approaches can be seen for an example set of curves in Figure~\ref{fig:fit}. Unmodified FPCA is ill-suited to account for sources of horizontal variation, like shift warping, as its eigenfunctions and their scores are geared towards representing vertical variation. In the presence of this horizontal variation, the estimated FPC scores then tend to over- or underestimate the actual amplitude variation, especially near the peaks, as seen in the left side of Figure~\ref{fig:fit}. By accounting for component warping with XCR however, the burden of modeling time domain variation is lifted from FPCA, which can then focus on amplitude variability without the confounding phase noise. Another simulation study characterizing finite sample performance at more noise levels can be found in the Supplement.

\section{Theoretical Results}

For bivariate Cross-Component Registration, a key finding is that the centered process
\begin{equation*}
Z_n(\tau)=\sqrt{n}\{L_n(\tau)-\Lambda(\tau)\}
\end{equation*}
converges weakly to a Gaussian limit process $Z(\tau)$, where $L_n, \Lambda$ are as 
in 
(\ref{eq-2.1.2}), (\ref{eq-2.1.3}). The details of this result are shown in Lemma 1 of the Supplement. To show weak convergence of the pairwise estimate $\widehat{\tau}$ as defined in (\ref{eq-2.1.4}), we require the following assumptions on $\Lambda$.
\noindent
\begin{enumerate}[label=(P\arabic*), align=left]
	\setcounter{enumi}{0}
	\item For any $\varepsilon>0,~
	\underset{\tau:d(\tau,\tau_0)>\varepsilon}{\inf} \Lambda(\tau) < \Lambda(\tau_0)$.
	\item  There exists $\eta>0$, $C>0$ and $\beta>1$, such that, when $d(\tau,\tau_0)<\eta$, we have   \label{last-item}
	\begin{equation*}
	\Lambda(\tau)-\Lambda(\tau_0)\geq C d(\tau,\tau_0)^\beta.
	\end{equation*}
\end{enumerate}
Assumption (P1) ensures that there exists a well-defined minimum, and assumption (P2) describes the local curvature of $\Lambda$ at the true minimum $\tau_0$, compare, e.g., \cite{mull:16:2}.
We also require the following  assumptions for the observed random processes.
\noindent
\begin{enumerate}[label=(A\arabic*), align=left]
	\setcounter{enumi}{0}
	\item  $X_{j}(t)$ is continuously twice differentiable for~$j=1,\ldots,p$,
	\item $E\{ \int_\mathcal{I} X_{j}^4(t)dt\}<\infty, $~for~$j=1,\ldots,p$,
	\item $E\{\int_\mathcal{I} X_{j}^{'4}(t)dt\}<\infty,$~for~$j=1,\ldots,p$.
\end{enumerate}

These assumptions are standard in the literature. They were, for example, previously stipulated  in \cite{hall:07:1} and  enable  us to obtain asymptotic covariance matrices for our estimates and to derive some crucial bounds.

{\Theorem
		In the bivariate case, under assumptions (P1)-(P2), and (A1)-(A3), we have
	\begin{equation*}
	\widehat{\tau}-\tau_0=O_p(n^{-1/{2(\beta-1)}}).
	\end{equation*}
	In particular, when $\beta=2$, the sequence $\sqrt{n}(\widehat{\tau}-\tau_0)$ is asymptotically normal with mean zero and variance  $V=4\int_{\mathcal{I}}E[\{X_{1}(t)-X_{2}(t-\tau_0)\}X^{'}_{2}(t-\tau_0)]^2dt/\{\Lambda^{''}(\tau_0)\}^2$\linebreak
	where $\Lambda(\tau)=E\int_{\mathcal{I}} \{X_{1}(t)-X_{2}(t-\tau)\}^2dt$.
}

The proof is in the appendix and utilizes results for $M$-estimators \citep{jain:75, well:96, vdv:98}. We note  that when the local geometry around the minimum has a quadratic curvature, i.e. when $\beta=2$, one obtains the  parametric rate $n^{1/2}$.

Our main result for general Cross-Component Registration concerns the rate of convergence of the estimated global shift vector and its asymptotic distribution, as follows.
{\Theorem In the general case, under assumptions (P1)-(P2) and (A1)-(A3) % imposed on the component processes $\{X_j(t)\}_{j=1}^p$,
\begin{equation*}
\widehat{\bm{\theta}}-\bm{\theta}_{0}=O_p(n^{-1/{2(\beta-1)}}).
\end{equation*}
In particular, when $\beta=2$, the sequence $\sqrt{n}(\widehat{\bm{\theta}}-\bm{\theta}_0)$ is asymptotically normal with mean zero and covariance matrix  $$\bm{\Sigma}_p=\dfrac{1}{p^2}\bm{A}^T
\left[	\begin{array}{cc}
\bm{V}^{-1}_{\bm\tau_0}E\left(\nabla \bm{m}_{\bm\tau_0}\nabla \bm{m}_{\bm\tau_0}^T\right)\bm{V}^{-1}_{\bm\tau_0} & 0\\
0&0
\end{array}\right]\bm{A},$$ where 
$\bm{m}_{\tau_0}=\{L_n(\tau_{12}),L_n(\tau_{13}),...,L_n(\tau_{(p-1)p})\}^T$ and $\bm{V}_{\tau_0}$ is the Hessian of $\bm{\Lambda}(\bm\tau)=E(\bm{m}_{\bm\tau})$ evaluated at $\bm\tau_0$.}

\section{Concluding Remarks}
Cross-component registration seeks to address mutual component time warping that is often an issue for multivariate functional data arising from longitudinal studies in the biosciences. This issue does not manifest itself for univariate functional data.  By focusing on time warping across components, and not on the traditional time warping between individual subjects, we are able to estimate population-wide time shift parameters with fast parametric rates of convergence and obtain a limit distribution under suitable assumptions.

This new cross-component time warping approach leads to insights about the relative timings of the component processes, which is of interest for the analysis of growth data and also other multivariate longitudinal data. After cross-component shift warps have been identified and incorporated into the model, common methods such as functional principal component analysis for multivariate processes can be expected to lead to more meaningful outputs and the resulting principal component component scores can be used for subsequent downstream analysis. The identification and estimation of the underlying latent process may also lead to a more parsimonious representation and is of interest in itself. 

There are limitations of the framework we have established here. While the shift-warping model we develop in this paper is appropriate for certain applications such as the Z\"urich Longitudinal Growth Study, the cross-component warping phenomena need not be restricted to shifts in general and may emerge in the form of non-linear distortions among components. Using the shift-warping methodology in such a situation may or may not yield satisfactory results, depending on the nature of the actual time warping. If the warp has a simple structure, a shift parameter may be a sufficient and parsimonious way to discover and approximate the component time relations, especially for practitioners who seek clear and concise interpretations. However, the situation for more pronounced or complicated warps is less auspicious. When the data at hand exhibit complex component warping beyond shifts, a more flexible warping paradigm should be adopted. The nonlinearity of such cross-component distortions may suggest that such problems warrant an alternative metric to the $\mathcal{L}^2$-norm.

In spite of this, we argue that the limitations of a shift-warping model are not necessarily tied to the general idea of cross-component registration which we have presented here. While in this paper we have used a shift-warping model to introduce the notion of cross-component registration, one can imagine more flexible extensions.  The study of nonlinear warping models in cross-component registration is left for future research. Other potential directions of interest concern alternative representations of the cross-component warping problem.

%%%%%% include this section if you wish to acknowledge people,
%%%%%% grant support, etc.

%\section*{Acknowledgements}
%We wish to thank two referees and an Associate Editor for constructive comments. This research was supported by an NIH grant and 
%NSF grants DMS-1712864 and DMS-2014626.
%\section*{Data Availability Statement}
%The data used in the application section are proprietary to the Z\"urich University Children's Hospital and therefore not shared.

%%%%%% include this section only if your manuscript refers to supplementary
%%%%%% materials -- see Instructions for Authors at
%%%%%% http://www.tibs.org/biometrics

%\bibliographystyle{} \bibliography{xc-ref-2.bib}

\references
\label{lastpage}

\end{document}